\documentstyle[11pt,paspconf,epsf]{article}

\begin{document}

\title{The PG X-Ray QSO Sample:
     Links among X-ray, UV \& Optical Spectra}

\author{Beverley J. Wills\altaffilmark{1}, M. S. Brotherton\altaffilmark{2},
A. Laor\altaffilmark{3}, D. Wills\altaffilmark{1}, B. J. Wilkes\altaffilmark{4},
and G. J. Ferland\altaffilmark{5}}
%


\altaffiltext{1}{McDonald Observatory \& Astronomy Department, University of
Texas at Austin, TX, USA}
\altaffiltext{2}{Institute of Geophysics \& Planetary Physics, Lawrence 
Livermore National Laboratory, Livermore, CA, USA}
\altaffiltext{3}{Department of Physics, Technion, Israel Institute of 
Technology, Haifa 32000, Israel}
\altaffiltext{4}{Center for Astrophysics, 60 Garden Street, Cambridge MA 02138,
USA}
\altaffiltext{5}{University of Kentucky, Department of Physics and Astronomy, 
Lexington, KY 40506}



\begin{abstract}
A unique, essentially complete sample of 22 QSOs, with high quality soft X-ray
 spectra
from ROSAT, as well as HST and optical spectrophotometry from below 
Ly$\alpha$\ to above H$\alpha$, is being used to investigate the 
relationships among the ionizing continuum and the optical and UV continuum,
emission and absorption lines.  Here we present a first analysis showing 
that optical `Eigenvector 1' linking steeper soft X-ray spectra with
increasing optical Fe\,II strength, decreasing [O\,III]\,$\lambda$5007 
emission, and narrower BLR H$\beta$\ emission, extends to the UV emission lines,
and is manifested by weaker C\,IV\,$\lambda$1549 emission, stronger
Si\,III]\,$\lambda$1892/C\,III]\,$\lambda$1909 ratio, and narrower 
C\,III]\,$\lambda$1909 emission.  Steeper soft X-ray spectra have been 
linked to higher L/L$_{\rm Edd}$\ ratios, thus apparently linking BLR
densities, high and low ionization gas, and kinematics, to the accretion
process.
\end{abstract}




\section{Introduction}

In principle, relationships between QSOs' EUV continuum and the emission-line
gas that it ionizes should give us clues to any relationship between
accretion power and the physical conditions and kinematics of accreting or
outflowing material within $\sim$ 1 pc, hence clues to the mechanism of the
central engine.  Several studies have shown that the soft X-ray spectrum is
related to the emission line spectrum: steeper X-ray spectra are associated with
stronger optical Fe\,II (BLR) emission, narrower (BLR) H$\beta$, weaker 
[OIII]\,$\lambda$5007 (NLR) emission (Boroson \& Green 1992; Grupe 1996; Grupe
et al. 1998;  Forster 1995; Laor et al. 1994, 1997; Corbin 1993).  One of the 
greatest sources of variation from one
spectrum to another can be represented as a linear combination of these
observables -- the so-called ``Eigenvector 1'' of principal component
analyses.  Its underlying physical cause is unknown, but an understanding
seems likely to hold a clue to accretion conditions, and to the energy budget
problem -- in particular the too-great strength of low-ionization emission
lines such as Fe\,II (Netzer 1985).

Laor et al. (1997) have investigated the soft X-ray and 
H$\beta$--[OIII]\,$\lambda$5007 region for a complete sample of all 23 QSOs from
the PG UV-excess survey (Schmidt \& Green 1983) with $z < 0.4$,
and low Galactic absorption (N$_{HI} < 1.9 \times 10^{20}$\,cm$^{-2}$), 
discovering strong Eigenvector 1 relationships in this sample.  The low 
redshift ensures detection of the soft X-ray emission down to the lowest
possible rest frame energy (typically 0.2~keV), which is
redshifted into the unobservable soft X-ray region ($<0.15 keV$) in higher
redshift quasars. The low Galactic absorbing column, and accurate 21~cm
measurements
of this column for all objects, ensure small, accurate corrections for 
ultraviolet and soft X-ray absorption.

This sample is ideal for extending this study into the ultraviolet,
where the highest energy continuum and important UV diagnostic lines can be
measured with minimal confusion from intergalactic absorption lines.  Thus we
have obtained HST FOS spectrophotometry from wavelengths below Ly$\alpha$ to
beyond the atmospheric cut-off, and McDonald Observatory spectrophotometry
from the atmospheric
cut-off to beyond H$\alpha$.  Instrumental resolutions range from 230 -- 350
km s$^{-1}$\,(FWHM).  Here we present highlights of a first look at our own and
archival HST spectra and the X-ray and optical measurements presented by Laor
et al. (1997), and Boroson \& Green (1992).

\section{Spectral Measurements, and Correlations}

We have measured strengths, ratios and widths (FWHM) for the following
emission lines: Ly$\alpha$ with
N\,V\,$\lambda$1240 removed, C\,IV\,$\lambda$1549 with N\,IV\,$\lambda$1486,
He\,II\,$\lambda$1640 and [O\,III]\,$\lambda$1663 removed, and we have deblended
Si\,III]\,$\lambda$1892 and C\,III]\,$\lambda$1909.   In most cases it was
possible to define a `rest frame' wavelength scale referred to 
[O\,III]\,$\lambda$5007 in our McDonald spectra.  Generally Fe\,III does not
contribute much to the $\lambda$1909 blend.  Evidence for this is that the
wavelength of the peak corresponds to within 0.5-1\AA\ rms of the expected
wavelength of C\,III]\,$\lambda$1909.
An exception is Mkn 478, where Fe\,III is a clear contributor.
The greatest uncertainties in line measurements arise from uncertainties in
continuum placement, and in removal of associated and Galactic interstellar
absorption.  Details will be presented by Wills et al. (1998b).

Table 1 presents a few of the correlation results.
Eigenvector 1 observables, given in the first column, are correlated with
important emission line parameters of the ultraviolet spectrum, given across
the top of the table.
Eigenvector 1 variables are chosen so as to correlate positively with X-ray
spectral index $\alpha_x$ (F$_{\nu} \propto \nu^{-\alpha_x}$).
Correlation coefficients are generally Pearson coefficients using line ratios,
and logarithms of equivalents widths and FWHMs.  Spearman rank correlations give
similar results.  The two-tailed significance levels are given at the end of
the table.  We note that a large fraction of our observationally-independent
parameters are correlated.  This means that the significance of an individual
correlation is not much affected by the fact that we attempted a large number
of correlations.  Figure\,1 plots some of the correlations of Table\,1,
the four columns representing
Eigenvector 1 observables: the steepness of the X-ray spectrum, the strength of
Fe\,II (optical), the strength of NLR emission
([O\,III]\,$\lambda$5007), and the width of the broad H$\beta$ line.
In Figure 2 we show the intensity ratio Si\,III]/C\,III] plotted against 
Fe\,II/H$\beta$
and Ly$\alpha$/C\,IV.

\begin{table}
\caption{`Eigenvector 1' Correlations.}
\label{tbl-1}
\begin{center}
\begin{tabular}{c@{\qquad}c@{\qquad}c@{\qquad}c@{\qquad}c@{\qquad}c@{\qquad}c}
\tableline
{\small Eigenvector\,1} & \multicolumn{6}{c}{\small UV Parameters}\\
{\small Parameters} & $\frac{{\rm Ly}\alpha}{{\rm C\,IV}}$ & 
             $\frac{\rm O\,VI}{\rm C\,IV}$ & $\frac{\rm C\,III]}{\rm C\,IV}$ & 
              {\scriptsize EW(C\,IV)} &
           $\frac{\rm Si\,III]}{\rm C\,III]}$ & {\scriptsize FWHM (C\,III])} \\
\\
\tableline
\\
{\scriptsize $\alpha_{x}$}    & 0.78 & 0.42 & $\cdots$ & $-$0.67 & $\cdots$ &
 $-$0.59 \\
{\scriptsize EW (Fe\,II)}       & 0.69 & 0.70 & 0.53 & $-$0.68 & 0.72 &
 $-$0.52 \\
$\frac{{\rm Fe\,II}}{{\rm H}\beta}$  & 0.80 & 0.85 & 0.54 & $-$0.46 & 0.89 &
 $-$0.63 \\
$\frac{{\rm Fe\,II}}{{\rm [O\,III]}}$   & 0.71 & 0.59 & 0.63 & $-$0.66 & 0.28 &
 $-$0.37 \\
$\frac{{\rm H}\beta}{{\rm [O\,III]}}$ & 0.58 & 0.40 & 0.56 & $-$0.59 & 0.56 &
 $-$0.47 \\
$\frac{1}{{\rm EW [O\,III]}}$  & 0.63 & 0.40 & 0.60 & $-$0.64 & 0.60 &
 $\cdots$ \\
$\frac{1}{{\rm FWHM (H}\beta)}$ & 0.68 & 0.65 & 0.53 & $-$0.69 & 0.56 &
 $-$0.78 \\
\end{tabular}
\end{center}
\tablenotetext{}{For 22 QSOs, a correlation coefficient of
0.4 corresponds to 1 chance in 15 of arising from uncorrelated variables 
(two tailed),
0.5 corresponds to 1 chance in 50, 
0.6 to 1 chance in 300, and 
0.7 to 1 chance in 2000 of arising from uncorrelated variables.\\
}
\end{table}
\noindent

\begin{figure}
\plotfiddle{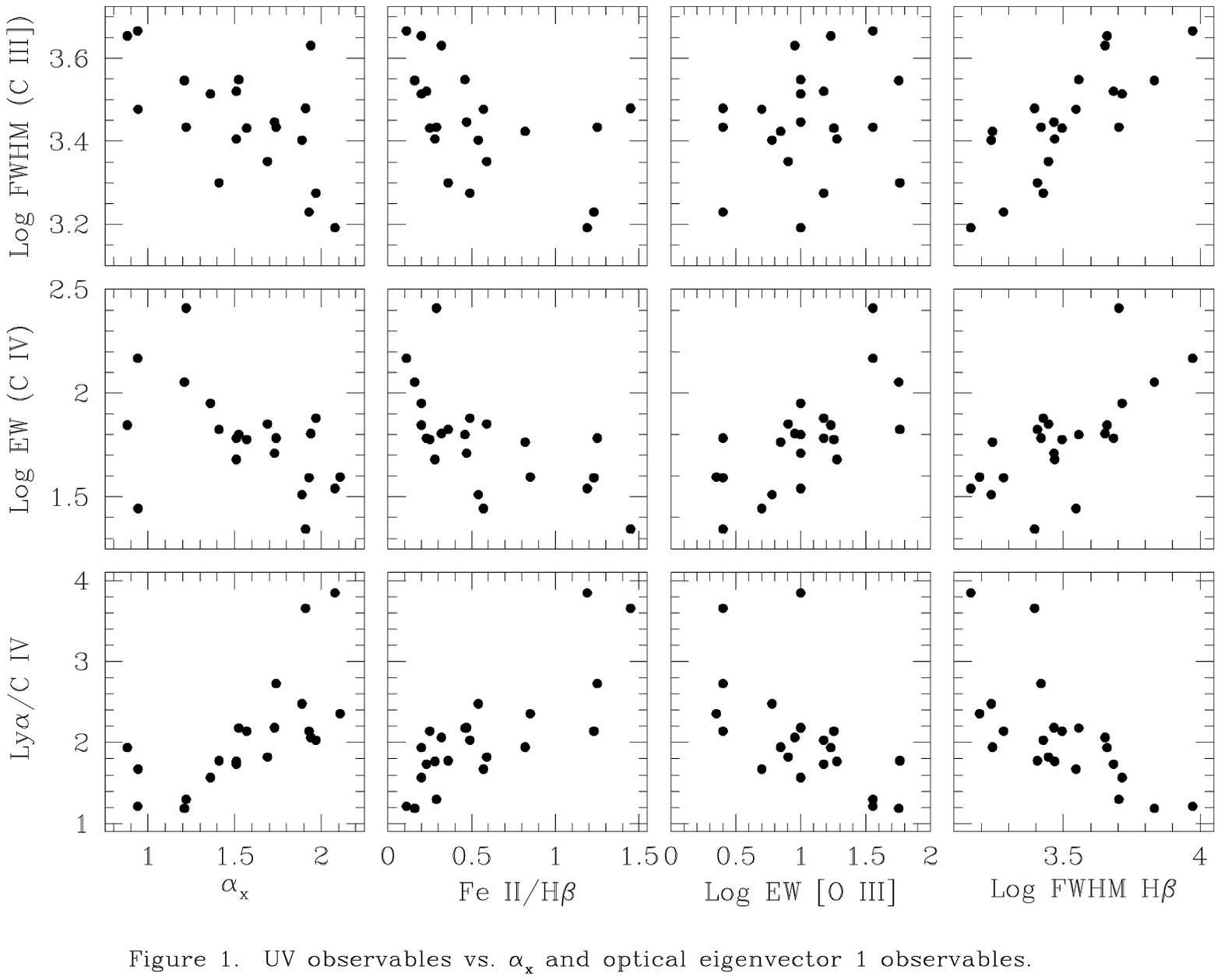}{21cm}{90.}{100}{100}{350.}{0.}
\end{figure}

\setcounter{figure}{1}

\begin{figure}
\plotone{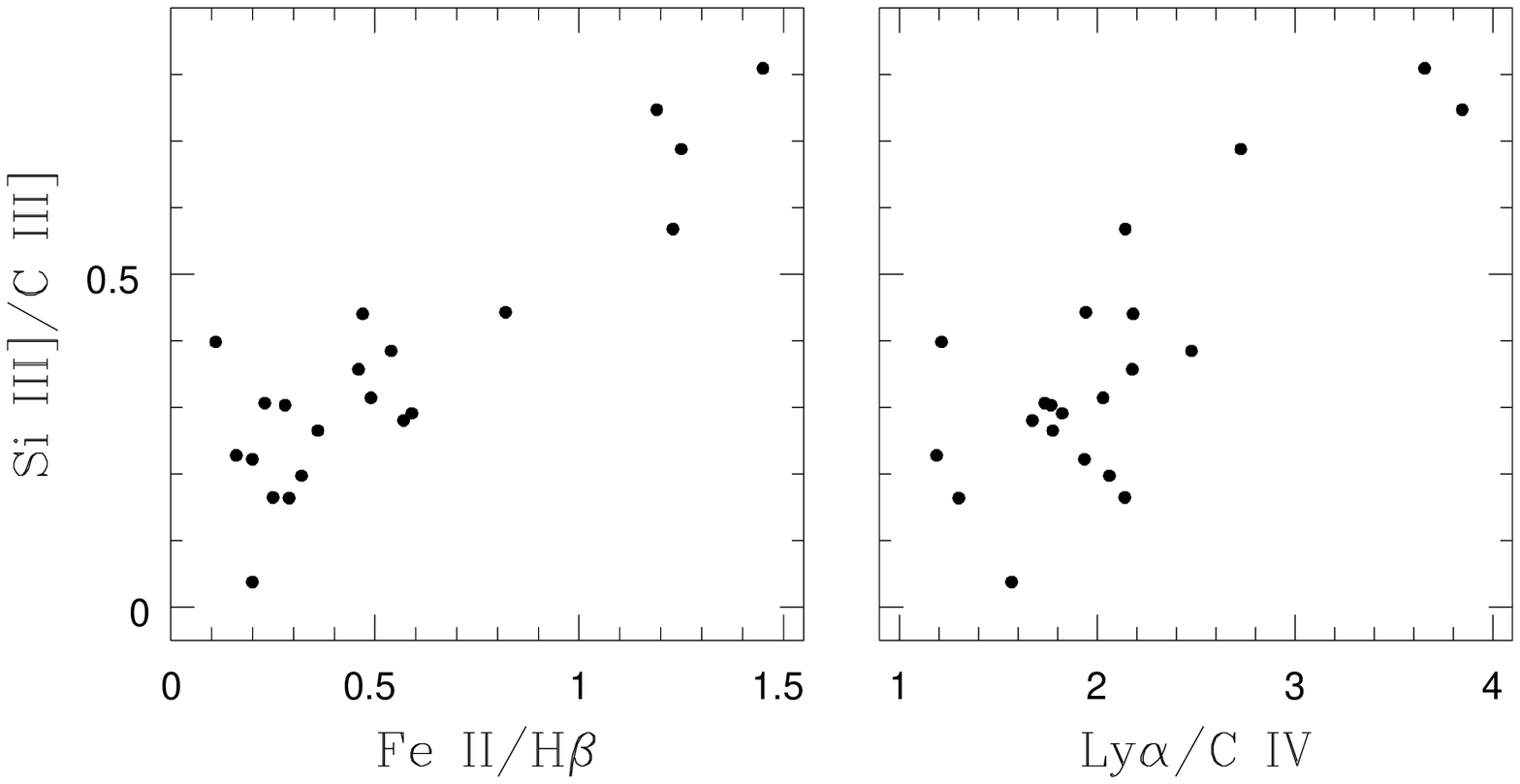}
\caption{Correlations with the possible density indicator,
Si\,III]\,$\lambda$1892/C\,III]\,$\lambda$1909.}
\end{figure}

\section{Results and Discussion}

Table 1 shows that line ratios involving C\,IV strength, including the 
EW (C\,IV),
correlate significantly with nearly all Eigenvector 1 observables in the
sense that C\,IV strength anticorrelates with steep soft X-ray spectrum, and
strong Fe\,II, and correlates positively with [O\,III] strength and 
FWHM (H$\beta$).
We note that correlations of Ly$\alpha$/C\,IV and EW (C\,IV) with $\alpha_x$
are in the same sense as found by Wang et al. (1998) for a large, heterogeneous
sample. 
Our result suggests that Eigenvector 1 is correlated with strengths of lines
from low ionization transitions (Fe\,II) and anti-correlated with higher
ionization transitions ([O\,III],
C\,IV, O\,VI\,$\lambda$1034).  If investigation of species over a wider range of 
ionization confirms this, one possibile explanation may be along the lines 
hinted at
by Boroson \& Green (1992), and investigated more quantitatively by
Brotherton (1996, 1998):
the [O\,III] may be produced by ionizing photons that reach the NLR after
penetrating the BLR.  Higher BLR optical depths  will result in stronger
emission from lower ionization lines (Mg\,II\,$\lambda$2798,
Fe\,II (optical)), hence reducing the ionizing flux to the NLR beyond.  
A more highly ionized BLR may produce stronger
O\,VI\,$\lambda$1034, C\,IV\,$\lambda$1549, and lower column densities
of Mg$^+$ and Fe$^+$, allowing greater ionizing flux to reach the NLR.
For example, in a BLR consisting of clouds with a distribution of optical
depths, increasing ionizing flux could increase the ratio of optically thin
to optically thick clouds.

The ratio Si\,III]\,$\lambda$1892/C\,III]\,$\lambda$1909 is strongly
correlated with Eigenvector 1 in the sense of a positive correlation with
Fe\,II (optical), and anticorrelations with [O\,III] strength and 
FWHM (H$\beta$).  It is also correlated with Ly$\alpha$/C\,IV (Fig. 2).
These correlations are stronger than the positive (negative) correlations
of Si\,III] strength (C\,III]) alone with Fe\,II/H$\beta$ and Ly$\alpha$/C\,IV.
The similar (but not identical) ionization potentials of Si$^{++}$\ and 
C$^{++}$\ and photoionization models suggest that these ions are largely
cospatial in the BLR.  This line ratio is sensitive to density, with
n$_{crit} \sim 10^{11}$~cm$^{-3}$\ for the Si\,III]\,$\lambda$1892 upper level
and n$_{crit} \sim 10^{9}$~cm$^{-3}$\ for C\,III]\,$\lambda$1909.  Thus
Eigenvector 1 appears strongly related to density, in the sense that the BLR
high density gas contributes most when Fe\,II (optical) is strongest.

As found in other samples, FWHM (C\,III]) correlates well with FWHM 
H$\beta$,
suggesting that these are from the same kinematic region.  Similarly, there is
a good correlation between FWHM (Ly$\alpha$) and FWHM (C\,IV), but
spanning a smaller range of FWHMs.  These correlations, together with weaker
correlations of FWHM Ly$\alpha$\ with FWHM (C\,III]) or FWHM H$\beta$, and
weaker correlations of FWHM (C\,IV) with FWHM (C\,III]) or FWHM H$\beta$\
suggest different kinematic origins for low and high ionization gas.

Overall, we were somewhat amazed at the strong correlations present in our
small sample, and conclude that small, but carefully-defined, homogeneous
samples with high quality spectra can yield significant information on the
underlying physics.
We suggest that Eigenvector 1, in the sense of increasingly steep soft 
X-ray spectrum, increasing Fe\,II (optical) strength, decreasing NLR emission,
and narrower (BLR) H$\beta$, represents increasing dominance of low-ionization
gas, and high densities, as well as decreasing dominance of high-ionization
emission.   The interpretation of narrower H$\beta$\ lines in terms
of higher Eddington ratios, and the correlation with steeper soft X-ray spectra,
links the accretion process
to density and ionization state of the surrounding gas (Wandel, Laor, these
Proceedings).

Further results, including measurements of other lines 
(e.g., N\,V\,$\lambda$1240, $\lambda$1400, 
Al\,III\,$\lambda$1860), will be presented by Wills et al. (1988a, b).

\acknowledgments

We gratefully acknowledge the help of the following people:
C. D. Keyes \& A. Roman of STScI, M. Dahlem (now of ESTEC),
Z. Shang, D. R. Doss, J. Martin, M. Villareal, M. Ward, D. Otoupal,
E. Green, D. Crook, \& M. Cornell of McDonald Observatory \& the 
University of Texas Astronomy Department.  This research is supported
by NASA through LTSA grant number NAG5-3431 (B.J.W.) and grant number
GO-06781 from the Space Telescope Science Institute, which is operated by
the Association of Universities for Research in Astronomy, Inc., under
NASA contract NAS5-26555.


%
%

\begin{question}{Dr.\ Binette}
Have you looked at other density indicators besides the ratio
Si\,III]\,$\lambda$1892/C\,III]\,$\lambda$1909?
\end{question}
\begin{answer}{Dr.\ Wills}
Not yet, but this will be possible (see Ferland 1998, 
Quasars as Standard Candles for Cosmology, ASP Conference Series).
\end{answer}
\begin{question}{Dr.\ Dultzin}
We find the same results as you have, and reported them in
Dultzin-Hacyan, D. 1997, `Emission Lines in Active Galaxies: New
Methods \& Techniques', ASP Conference Series, 113, 262.
\end{question}
\begin{answer}{Dr.\ Wills}
After the meeting I have checked this.  While your reported 
anticorrelation between EW (C\,IV) and Fe\,II (optical) strength
may be the same as what we find, I'm not sure.  Your EW (C\,IV)
apparently was measured in a way rather different from the 
conventional one (Marziani et al. 1996, \apjs, 104, 37)
and the sample you used was very heterogenous, covering a wide
range in luminosity.  Our EW (C\,IV) measurements include the total
line flux, but with N\,IV\,$\lambda$1486, He\,II\,$\lambda$1640, and 
O\,III]\,$\lambda$1663 removed.

\end{answer}
\end{document}